\documentclass[12pt]{article}
\usepackage{mint,psfig,mycite}

\title{
\vspace{-1cm}
\begin{flushright}
\normalsize 
HUPD-0201 \\
\end{flushright}
\vspace{4cm}
{\Large \bf Phenomenological approach to symmetry breaking pattern of 
 democratic mass matrix}}

\author{\hspace{-0.8cm}
Junpei Harada\footnote{e-mail: harada@theo.phys.sci.hiroshima-u.ac.jp} 
        \\[1.2em]
	{\it Department of Physics, Hiroshima University,
		Hiroshima 739-8526, Japan} }

\date{}

\begin{document}
\baselineskip 0.6cm

\maketitle

\begin{abstract}
 We investigate the symmetry breaking pattern of the democratic mass
 matrix model, which leads to the small flavor mixing in quark sector
 and bi-large mixing in lepton sector. We present the symmetry breaking
 matrices in quark sector which are determined by alternative ways
 instead of conventional ansatz. These matrices might be useful for
 understanding the origin of democratic symmetry and its breaking.
\end{abstract}

\newpage

\section{Introduction}
\label{sec:Introduction}
 One of the most important problems in particle physics is to understand
 the flavor structure of quarks and leptons. Since the feature of the 
 flavor structure appear in the quark-lepton mass matrix, it is
 important to investigate the quark-lepton mass matrix for understanding
 the flavor structure.

 Various phenomenological mass matrix models of quarks and leptons are
 proposed. One of the most attractive mass matrix models is democratic
 mass matrix model~\cite{HHW78}, which has $S_{3L}\times S_{3R}$
 permutaion symmetry (democratic symmetry). This symmetry means that
 right-handed fermion and left-handed fermion independently have the 
 permutaion symmetry among three generations. This democratic mass
 matrix model induces the small mixing angles in quark sector and
 bi-large mixing angles in lepton sector~\cite{SK98}\cite{SNO02}. 
 The democratic mass matrix model also induces small mixing angle
 $U_{e3}$ in lepton sector, which is consistent with the experimental
 bound from CHOOZ reactor experiment~\cite{CHO99}. 
 Thus the democratic mass matrix model is very successful phenomenologically.

 Inspite of these successes, however, the democratic mass matrix induces
 the massless quarks and charged leptons of the first and the second
 generations. Then Cabbibo-Kobayashi-Maskawa (CKM) quark flavor mixing
 matrix is exactly equals to a unit matrix in the democratic symmetry
 limit. To aquire realistic fermion masses and mixing angles, 
 democratic symmetry must be broken and small symmetry breaking terms
 are required. If we know the mechanism of democratic
 symmetry  breaking, we can predict the quark-lepton mass spectrum and
 all mixing angles. Unfortunately, however, such symmetry breaking
 mechanism is not known. In the literature, some works have assumed
 these small symmetry breaking terms and given some predictions.
 This approach has been known very successful for quark
 sector~\cite{K83}\cite{KM88}
 and lepton sector~\cite{FTY98}\cite{FZ98}\cite{FHY02}. 
 However, there are possibilities
 that other ansatz can also account for the observed masses and mixing
 angles. Then it is very important to understand the mechanism of 
 democratic symmetry breaking.

 In this letter, we present the symmetry breaking terms which are
 determined by alternative ways instead of conventional ansatz. These symmetry
 breaking terms might be useful for understanding the origin of democratic
 symmetry and the mechanism of its breaking. Although we are especially
 interested in the lepton sector, we investigate the symmetry breaking
 terms of quark sector in this letter. 
 The reasons are as follows. First, the experimental information
 of quark sector is much rather than that of lepton sector. Second,
 since both up-type quarks and down-type quarks are Dirac particles, they
 are treated in an equal footing in the democratic mass matrix
 model. Third, we expect that the symmetry breaking pattern of down
 quarks can be applied for charged leptons.

 This letter is organized as follows.
 In sec.~2 we briefly review the dmocratic mass matrix model. In sec.~3
 we present the symmetry breaking terms of quark sector. In sec.~4 we
 comment on lepton sector. Conclusion is denoted in sec.~5.

\section{Democratic Mass Matrix}
\label{sec:Democratic}

 We briefly review the democratic mass matrix model. At first, we
 concentrate on quark sector. The democratic mass matrix has
 $S_{3L}\times S_{3R}$ permutation symmetry, i.e., 
\begin{eqnarray}
  M_u^{[0]} 
  = 
  \frac{m_t}{3}
  \left[
   \begin{array}{ccc}
     1 & 1 & 1 \\
     1 & 1 & 1 \\
     1 & 1 & 1 
   \end{array}
  \right]
  \,,
\end{eqnarray}
 where $m_t$ is top quark mass. The down quark mass matrix has the same
 structure. This matrix is a unique representation of the $S_{3L}\times
 S_{3R}$ symmetric matrix. This matrix is diagnalized as  
\begin{eqnarray}
  V_u^{[0]\dagger} M_u^{[0]} V_u^{[0]}
  =
  M_u^{[0]diag}
  = 
  diag(0,0,m_t)
  \,.
\end{eqnarray}
 The unitary matrix $V_u^{[0]}$ that diagnalizes the democratic mass matrix
 $M_u^{[0]}$ is 
\begin{eqnarray}
  V_u^{[0]} 
  = 
  \left[
   \begin{array}{ccc}
     1/\sqrt{2} & 1/\sqrt{6} & 1/\sqrt{3} \\
    -1/\sqrt{2} & 1/\sqrt{6} & 1/\sqrt{3} \\
     0          &-2/\sqrt{6} & 1/\sqrt{3}
   \end{array}
  \right]
  \,.
\end{eqnarray}
 Since there is no difference between up quark mass matrix and down
 quark mass matrix(except for $m_t \not= m_b$), unitary matrices that
 diagonalizes mass matrix of up quarks and one of down quarks are same each
 other, i.e., $V_u^{[0]} = V_d^{[0]}$. Then
 Cabbibo-Kobayashi-Maskawa(CKM) quark flavor mixing matrix is predicted
 as 
\begin{eqnarray}
  V_{CKM}^{[0]}
  = 
  V_u^{[0]\dagger}V_d^{[0]} 
  = 
  diag(1,1,1)
  \,.
\end{eqnarray}

 From experimental observations, it is well known that 
\begin{eqnarray}
  M_q^{diag} 
   \simeq 
  diag(\epsilon_q^4,\epsilon_q^2,1)m_{t/b}
  \,,
\end{eqnarray}
 where $q = u$ or $d$, and CKM mixing matrix in Wolfenstein
 parametrization is given by
\begin{eqnarray}
  V_{CKM} 
  =
  \left[
   \begin{array}{ccc}
    1-\frac{\lambda^2}{2} & \lambda               & 0 \\
    -\lambda              & 1-\frac{\lambda^2}{2} & A\lambda^2 \\
    0                     & -A\lambda^2           & 1 
   \end{array}
  \right]
  + 
  O(\lambda^3)
  \,.
\end{eqnarray}
 Thus, in quark sector, the predictions of the democratic mass matrix model
 are completely consistent with experimental data with approprite
 approximations, i.e., 
\begin{eqnarray}
  M_u^{diag} 
 & = &
  diag(0,0,m_t) 
  + 
  O(\epsilon_u^2)
  \,,\\
  M_d^{diag} 
 & = &
  diag(0,0,m_b) 
  + 
  O(\epsilon_d^2)
  \,,\\
  V_{CKM}
 & = &
  diag(1,1,1) 
  + 
  O(\lambda)
  \,.
\end{eqnarray}

 Let's turn to the lepton sector. Since the charged leptons are Dirac
 particles, the charged lepton mass matrix has the same structure of
 quark mass matrix.
 Assuming that the neutrinos are the Majorana type, there are two
 independent mass matrices that are invariant under $S_{3L}$ permutation
 symmetry~\cite{FTY98}.  Then democratic neutrino mass matrix is given
 as follows;
\begin{eqnarray}
  M_\nu^{[0]} 
  = 
  m_\nu
  \left[
   \begin{array}{ccc}
     1 & 0 & 0 \\
     0 & 1 & 0 \\
     0 & 0 & 1 
   \end{array}
  \right]
  +
  m_\nu  r
  \left[
   \begin{array}{ccc}
     0 & 1 & 1 \\
     1 & 0 & 1 \\
     1 & 1 & 0 
   \end{array}
  \right]
  \label{eq:neutrino-democratic}
  \,,
\end{eqnarray}
 where $r$ is arbitrary parameter. Here we take $r=0$, since it
 automatically leads to the bi-large mixing angles of solar and
 atmospheric neutrinos {\it and} small $U_{e3}$ 
 (We comment on this choice in sec.~4).
 Thus in the democratic mass matrix model, neutrino masses of three
 generations are degenerate. Then Maki-Nakagawa-Sakata(MNS) lepton
 flavor mixing matrix is given by 
\begin{eqnarray}
  U_{MNS}^{[0]} 
  = 
  V_l^{[0]\dagger}
  =
  \left[
   \begin{array}{ccc}
     1/\sqrt{2} &-1/\sqrt{2} & 0 \\
     1/\sqrt{6} & 1/\sqrt{6} &-2/\sqrt{6} \\
     1/\sqrt{3} & 1/\sqrt{3} & 1/\sqrt{3}
   \end{array}
  \right]
  \,,
\end{eqnarray}
 where $V_l^{[0]}$ is the unitary matrix that diagonalizes the
 charge lepton mass matrix. The predictions of mixing angles are 
\begin{eqnarray}
  \sin^2 2\theta_{sol} 
  = 
  1
  \,, \quad
  \sin^2 2\theta_{atm} 
  = 
  \frac{8}{9}
  \,, \quad
  U_{e3}
  =
  0
  \,.
  \label{eq:lepton-mixing}
\end{eqnarray}
 Thus, in lepton sector, the predictions of democratic mass matrix model
 are consistent with experimental data with approprite approximations.

 In summary, the democratic mass matrix model is phenomenologically very
 successful in both quark sector and lepton sector.

\section{Symmetry Breaking Matrix}
\label{sec:SB}

 In previous section, we saw the democratic mass matrix is good
 candidate phenomenologically. However, the democratic mass matrix
 predict the massless quarks and charged leptons of the first and the second
 generations. It also predicts that CKM mixing matrix equals to a unit matrix.
 To aquire realistic fermion masses and mixing angles, small symmetry
 breaking terms are required. In this section, we present such symmetry
 breaking terms. We concentrate on the quark sector as denoted in the
 introduction.

 We determine the symmetry breaking terms order by order. For example,
 $V_{ub}$ equals zero in this letter. 
 If one want to induce realistic $V_{ub}$, higher order correction terms
 are required. This point is crucial in our approach.

 At first, we consider the symmetry breaking terms which leads to
 realistic quark masses, while CKM mixing matrix remains a unit
 matrix (We consider the effects of deviation from a unit matrix later).
 In this letter, we assume that up quark mass matrix and down quark mass
 matrix are symmetric mass matrices because of democratic principle.
 We denote the diagonal mass matrix $M_q^{diag}$ as follows, 
\begin{eqnarray}
  M_q^{diag} 
 & \equiv &
  M_q^{[0]diag} +  M_q^{[1]diag} +  M_q^{[2]diag}
  \,, \\
 & = &
  diag(0,0,1)m_{t/b}
   +
  diag(0,\epsilon_q^2,0)m_{t/b}
   + 
  diag(\epsilon_q^4,0,0)m_{t/b}
  \,.
\end{eqnarray}
 Since the unitary matrix $V_q^{[0]}$ that diagonalizes the
 democratic mass matrix $M_q^{[0]}$ is already known, as long as CKM mixing
 matrix remains a unit matrix, we can {\it uniquely} determine the symmetry
 breaking terms 
 which lead to the massive quarks of the first and the second generations,
\begin{eqnarray}
  M_q
 & = &
  V_q^{[0]} M_q^{diag} V_q^{[0]\dagger}
  \,, \nonumber \\
 & = &
   \sum_{i=0}^2 V_q^{[0]} M_q^{[i]diag} V_q^{[0]\dagger}
  \,, \nonumber \\
 & \equiv &
  M_q^{[0]} +  M_q^{[1]} +  M_q^{[2]}
  \,, \nonumber \\
 & = &
  \frac{m_{t/b}}{3}
  \left(
   \left[
   \begin{array}{ccc}
     1 & 1 & 1 \\
     1 & 1 & 1 \\
     1 & 1 & 1 
   \end{array}
   \right]
   +
   \epsilon_q^2
   \left[
   \begin{array}{ccc}
     1/2 & 1/2 & -1 \\
     1/2 & 1/2 & -1 \\
     -1  & -1  &  2 
   \end{array}
   \right]
   +
   \epsilon_q^4
   \left[
   \begin{array}{ccc}
     3/2 &-3/2 & 0 \\
    -3/2 & 3/2 & 0 \\
     0   & 0   & 0 
   \end{array}
    \right]
  \right)
   \,. 
   \label{eq:real-mass}
\end{eqnarray}
 Note here that the first order symmetry breaking term, which leads to
 the charm/strange quark masses, automatically has $S_{2L}\times S_{2R}$
 permutaion symmetry~\cite{FZ00}.

 Next we consider the symmetry breaking terms which lead to realistic
 quark flavor mixing angles. Here we assume that $M_u$ and $M_d$ are the
 fuction of $\epsilon_u$ and $\epsilon_d$, respectively. 
 We take CKM mixing matrix in
 $O(\lambda^2)$ approximation, i.e., 
\begin{eqnarray}
  V_{CKM} 
  =
  \left[
   \begin{array}{ccc}
    1        & \lambda & 0 \\
    -\lambda & 1       & 0  \\
    0        & 0       & 1 
   \end{array}
  \right]
  + 
  O(\lambda^2)
  \,.
\end{eqnarray}
 Note here that $CP$ is conserved at this order since $CP$ phase exits in 
 $O(\lambda^3)$ terms in Wolfenstein parametrization. 
 We denote the uniraty matrix that diagonalizes up/down mass matrix
 as follows,
\begin{eqnarray}
  V_q 
  = 
  V_q^{[0]} + V_q^{[1]} + \cdots
  \,,
\end{eqnarray}
 where $V_q^{[1]}, \cdots$ are small correction terms which depend on
 $\epsilon_q$.
 Then CKM mixing matrix is given by 
\begin{eqnarray}
  V_{CKM}
  =
     V_u^{[0]\dagger}V_d^{[0]}
   + V_u^{[1]\dagger}V_d^{[0]}
   + V_u^{[0]\dagger}V_d^{[1]}
   + V_u^{[1]\dagger}V_d^{[1]}
   + \cdots
  \,. 
\end{eqnarray}
 The first term obviously equals to a unit matrix, which is zero-th
 order CKM mixing matrix $V_{CKM}^{[0]}$. Since the fourth term is
 higher order rather than the second and the third terms, we assume that the
 second and third terms represent $O(\lambda)$ terms in CKM matrix, 
\begin{eqnarray}
  V_{CKM}^{[1]}
  \equiv
  \left[
   \begin{array}{ccc}
    0        & \lambda & 0 \\
    -\lambda & 0       & 0  \\
    0        & 0       & 0 
   \end{array}
  \right]
  =
     V_u^{[1]\dagger}V_d^{[0]}
   + V_u^{[0]\dagger}V_d^{[1]}
  \,. 
  \label{eq:CKM1st}
\end{eqnarray}
 This equation should hold for arbitrary $\epsilon_q$ and 
 we can derive the first order correction term $V_q^{[1]}$
,
\begin{eqnarray}
  V_{q}^{[1]}
  =
  a_q
  \left[
   \begin{array}{ccc}
    -1/\sqrt{6} & 1/\sqrt{2} & 0 \\
    -1/\sqrt{6} &-1/\sqrt{2} & 0 \\
     2/\sqrt{6} & 0          & 0
   \end{array}
  \right]
  \,,
\end{eqnarray}
 where $\sin \theta_{12}^q \equiv \lambda = a_d - a_u$. Note here that
 parameter $a_q$ depends on $\epsilon_q$. Using this
 symmetry breaking terms of unitary matrix, we can determine the symmetry
 breaking mass matrix which induces $O(\lambda)$ terms in CKM mixing matrix,
\begin{eqnarray}
  M_q 
  & = &
  V_q M_q^{diag} V_q^{\dagger}
  \,, \nonumber \\
  & = &
  \sum_{i=0}^{2} V_q M_q^{[i]diag} V_q^{\dagger}
  \,, \nonumber \\
  & = &
  \sum_{i=0}^{2}
   \left( V_q^{[0]} M_q^{[i]diag} V_q^{[0]\dagger}  
        + V_q^{[1]} M_q^{[i]diag} V_q^{[0]\dagger}  
        + V_q^{[0]} M_q^{[i]diag} V_q^{[1]\dagger}  
        + \cdots
   \right)
  \,.
\end{eqnarray}
 The first term represents eq.~(\ref{eq:real-mass}). The second and
 third terms which induces $O(\lambda)$ terms in CKM mixing matrix are
\begin{eqnarray}
  \sum_{i=0}^{2}
   \left(
         V_q^{[1]} M_q^{[i]diag} V_q^{[0]\dagger}  
       + V_q^{[0]} M_q^{[i]diag} V_q^{[1]\dagger}  
   \right)
  = 
  \frac{m_{t/b}}{3}
  \sqrt{3} a_q 
  (\epsilon_q^2 - \epsilon_q^4)
  \left[
   \begin{array}{ccc}
    1 & 0 &-1 \\
    0 &-1 & 1 \\
   -1 & 1 & 0 
   \end{array}
  \right]
  \,.
\end{eqnarray}

 In summary, the democratic quark mass matrix with small symmetry
 breaking terms which lead to the realistic quark masses and CKM mixing
 matrix in $O(\lambda^2)$ approximation is given by 
\begin{eqnarray}
  M_q
 & = &
  \frac{m_{t/b}}{3}
  \left(
   \left[
   \begin{array}{ccc}
     1 & 1 & 1 \\
     1 & 1 & 1 \\
     1 & 1 & 1 
   \end{array}
   \right]
   +
   \epsilon_q^2
   \left[
   \begin{array}{ccc}
     1/2 & 1/2 & -1 \\
     1/2 & 1/2 & -1 \\
     -1  & -1  &  2 
   \end{array}
   \right]
   +
   \epsilon_q^4
   \left[
   \begin{array}{ccc}
     3/2 &-3/2 & 0 \\
    -3/2 & 3/2 & 0 \\
     0   & 0   & 0 
   \end{array}
    \right] 
  \right) \nonumber \\
  &&
  +
  \frac{m_{t/b}}{3}
  \left(
  \sqrt{3} a_q 
  (\epsilon_q^2 - \epsilon_q^4)
  \left[
   \begin{array}{ccc}
    1 & 0 &-1 \\
    0 &-1 & 1 \\
   -1 & 1 & 0 
   \end{array}
  \right]
  \right)
   \,. 
\end{eqnarray}
 This mass matrix is diagonalized as 
\begin{eqnarray}
  M_q^{diag} 
  = 
  V_q^\dagger M_q V_q
  =
  diag(\epsilon_q^4,\epsilon_q^2,1)m_{t/b}
  \,,
\end{eqnarray}
 where unitary matrix $V_q$ is 
\begin{eqnarray}
  V_q
  = 
  \left[
   \begin{array}{ccc}
     1/\sqrt{2} & 1/\sqrt{6} & 1/\sqrt{3} \\
    -1/\sqrt{2} & 1/\sqrt{6} & 1/\sqrt{3} \\
     0          &-2/\sqrt{6} & 1/\sqrt{3}
   \end{array}
  \right]
  +
  a_q
  \left[
   \begin{array}{ccc}
    -1/\sqrt{6} & 1/\sqrt{2} & 0 \\
    -1/\sqrt{6} &-1/\sqrt{2} & 0 \\
     2/\sqrt{6} & 0          & 0
   \end{array}
  \right]
  \,,
\end{eqnarray}
 where $\sin \theta_{12}^q \equiv \lambda = a_d - a_u$. 
 We summarize some assumptions we used ;
 1) Mass matrix of up quark and one of down quark are symmetric
 matrix. 2) Mass matrix $M_u$ and $M_d$ are the function of $\epsilon_u$
 and $\epsilon_d$, respectively. 3) Eq.~(\ref{eq:CKM1st})
 should hold for arbitrary $\epsilon_q$.

 It is straightforward to determine higher order corrections in our approach.

\section{Comment on Lepton sector}
\label{sec:MNS}

 Here we comment on lepton sector.
 In the democratic symmetry limit, the predicted lepton flavor mixing
 angles are given in eq.~(\ref{eq:lepton-mixing}). If the symmetry
 breaking terms of charged leptons are the same as the ones of down
 quarks\footnote{
 We expect the
 symmetry breaking terms of down quark mass matrix can be applied for
 the ones of charged leptons. This is one of the reasons why we
 concentrate on the quark sector in this letter.}, we can estimate the
 correction from zero-th order 
 prediction, i.e., eq.~(\ref{eq:lepton-mixing}). 
 Note here that
 neutrinos are symmetry limit. In this case, lepton mixing angles
 are predicted as follows\footnote{Here we assumed $a_d \gg a_u$ and
 this leads to $a_d \sim O(\lambda)$. This reason is as
 follows. Since the unitary matrix $V_q$ should become $V_q^{[0]}$ in the
 $\epsilon_q^2 \rightarrow 0$ limit, $a_q$ (parameter in the unitary
 matrix $V_q$) might be proportional to $(\epsilon_q^2)^x$ (where $x$ is
 positive unknown parameter). Thus $\lambda$ might be expressed by $\lambda =
 (\epsilon_d^2)^x a_d^\prime - (\epsilon_u^2)^x a_u^\prime$. 
 This relation is
 suitful for well-known phenomenological relation~\cite{T98}, $\lambda \sim
 \sqrt{m_d/m_s}$.
 }, 
\begin{eqnarray}
  \sin^2 2\theta_{sol} 
  \sim
  0.97
  \,, \quad
  \sin^2 2\theta_{atm} 
  \sim
  0.88
  \,, \quad
  U_{e3}
  \sim
  0.18
  \,.
\end{eqnarray}
 These predictions are slightly different from experimental data.
 This facts might indicate that we should consider $O(\lambda^2)$
 corrections in CKM mixing matrix and/or the off-diagonal elements {\it and}
 small symmetry breaking terms of neutrino sector.

 Then we present the estimation of the lepton flavor mixing angles 
 when we include $O(\lambda^2)$ corrections in CKM mixing matrix 
 (Note here that neutrinos are also symmetry limit.), 
\begin{eqnarray}
  \sin^2 2\theta_{sol} 
  \sim
  0.87
  \,, \quad
  \sin^2 2\theta_{atm} 
  \sim
  0.80
  \,, \quad
  U_{e3}
  \sim
  0.18
  \,.
\end{eqnarray}
 Note that there is no effects for $U_{e3}$ from $O(\lambda^2)$ corrections.
 Although $O(\lambda^3)$ terms in the unitary matrix $V_l$ can modify
 $U_{e3}$, this effects would be neumerically too small to be
 consistent with experimental upper bound $U_{e3} < 0.15$~\cite{CHO99}.
 Thus the off-diagonal elements and the symmetry breaking terms in
 neutrino sector might be very
 important to obtain the realistic lepton mixing angles. In other words,
 small $r$ parameter\footnote{
 It is interesting that higher dimensional GUTs
 model which leads this situation is recently proposed~\cite{WY02}.} in
 eq.~(\ref{eq:neutrino-democratic}) 
 {\it and} small symmetry breaking terms~\cite{FHY02} are required.

\section{Conclusion}
\label{sec:Conclusion}

 In this letter, we investigate the symmetry breaking pattern of the
 democratic mass matrix model. We present the symmetry breaking terms in
 quark sector which are determined by alternative ways instead of
 conventional ansatz. These symmetry breaking terms might be useful for
 understanding the origin of democratic symmetry and the mechanism of its
 breaking. We also estimate the lepton fravor mixing angles by assuming 
 that the symmetry breaking terms of charged leptons are the same as the
 ones of down quarks. The results indicate that to aquire realistic
 lepton mixing angles, the small
 off-diagonal elements {\it and} the symmetry breaking terms in neutrino mass
 matrix are important.

\section*{Acknowledgments}

We would like to thank T.~Inagaki, T.~Onogi, K.~Ohkura and M.~Tanimoto
for useful discussions.

\end{document}